\documentclass{ifacconf}

\usepackage{natbib,amsmath,amssymb}            
\usepackage{algorithm,algpseudocode,lscape}   
\usepackage[usenames,dvipsnames]{xcolor}    
\usepackage{graphicx}          
\usepackage{amsmath,amssymb,here,relsize}
\usepackage{graphicx,psfrag,color}
\usepackage{cases}
\usepackage{pstricks}
\usepackage[T1]{fontenc}

\newtheorem{ass}{\bf Assumption}

\newtheorem{com}{\bf Remark}

\newcommand{\bu}{\mathbf{u}}
\newcommand{\bw}{\mathbf{w}}
\begin{document}

\begin{frontmatter}

\title{On Adaptive Measurement Inclusion Rate In Real-Time Moving-Horizon Observers} 

\author[gipsa]{Mazen Alamir}

\address[gipsa]{University of Grenoble, Gipsa-lab, BP 46, Domaine Universitaire, 38400 Saint Martin d'H\`{e}res, France. (mazen.alamir@grenoble-inp.fr)}


\begin{abstract}
\noindent This paper investigates a self adaptation mechanism regarding the rate with which new measurements have to be incorporated in Moving-Horizon state estimation algorithms. This investigation can be viewed as the dual of the one proposed by the author in the context of real-time model predictive control. An illustrative example is provided in order to assess the relevance of the proposed updating rule. 
\end{abstract}

\end{frontmatter}

\section{Introduction}
\noindent Moving-Horizon Observers (MHO) are algorithms that involve repeated on-line optimization in order to update the estimated value of the state \citep{Michalska1995, Alamir99,Rao2003,Kuhl2011}. More precisely, the estimated state is the optimal solution of an optimization problem in which the cost function accounts for the interpretation of the past measurement (over some observation horizon) and the compatibility with the presumed model of the system. Periodically, the observation horizon is shifted in order to take into account new measurements while discarding older ones. Typically, the shifting period is taken equal to the measurement acquisition period which, in the early stage of MHO development was compatible with the assumption of instantaneous solvability of the optimization problem. \ \\ \ \\ 
Amazingly enough, this feature has never been revisited despite the recent advances in real-time implementation framework where the optimization is truncated before an optimal solution is reached. In such frameworks, the time that is available for the optimization is tightly related to the horizon shifting period. More precisely, the question is the following: 
\begin{quotation}
\noindent \sl How many iterations of the optimization algorithm should be executed before the cost function is updated by including new available measurements? 
\end{quotation}
The same question can be reformulated equivalently:\begin{quotation}
\noindent \sl How many new measurement instances should be acquired before the cost function is updated? (During how many periods should the optimizer work on the same unchanged problem before the latter is updated to account for the newly available measurements?)
\end{quotation}
To assess the relevance of the questions above, one can simply argue that it has never been proved that the commonly used answer is optimal in all circumstances. Yet, there is another, probably more convincing argument, which is that recent investigations (see \citep{Pavia2008,AlamirECC2013}) on the dual problem of Model Predictive Control (MPC) context clearly showed the existence of time varying, context dependent, optimal control updating period (which is the period during which the optimizer works on the same problem before the control defined by the updated parameter is applied to the system).\ \\ \ \\  In this paper, it is shown that a similar formulation, similar techniques and, under certain circumstances, similar conclusions can be obtained regarding the optimal  measurement inclusion rate problem.\ \\ \ \\ The paper is organized as follows: First, the problem is stated in section \ref{secpbstat} by showing that monitoring the measurement inclusion rate can be viewed as a discrete-time output regulation problem in which the control input is the number of iterations to be performed before new set of measurements is accounted for.  In section \ref{secupdatinglaw}, a gradient-based heuristic is proposed to solve this problem leading to an updating law for the measurement inclusion rate. An illustrative example is given in section \ref{secillust} in order to assess the efficiency of the proposed updating scheme and its ability to handle varying circumstances during the system's lifetime. Finally, section \ref{secconc} concludes the paper and gives hints for further investigation.  
\section{The Measurement Inclusion Rate As An Output Regulation Problem} \label{secpbstat} 
\noindent Let us consider dynamical systems that are governed by the following evolution law:
\begin{eqnarray}
x(k)&=&X(M,x(k-M),\bu(k)) \label{defdesyst} \\
y(k)&=&h(x(k),u(k)) \label{defdeyh} 
\end{eqnarray} 
where $x\in \mathbb{R}^{n}$ is the state vector, $M$ is some integer and where the notation $x(k)$ refers to the state at instant $k\tau$ for some sampling period $\tau$ that is supposed here (without loss of generality) equal to the measurement acquisition period.  $\bu(k)$ represents the sequence of some measured exogenous inputs such that:
\begin{eqnarray}
\bu(k) = \{u(k-M),\dots,u(k)\} \label{defdeubold} 
\end{eqnarray}  
Note that equation (\ref{defdesyst}) represents a multi-step state transition map that gives the value of the state vector starting from the initial value $x(k-M)$ and under the input sequence defined over the time interval $[k-M,k]$ by $\bu(k)$. In the sequel, the length of the sequence of inputs $\bu(k)$ is defined from the context so that the same notation $\bu(k)$ can be used to designate sequences of different lengths provided that the last input is $u(k)$. \ \\ 
\begin{figure}[H]
\begin{center}
\includegraphics[width=0.45\textwidth]{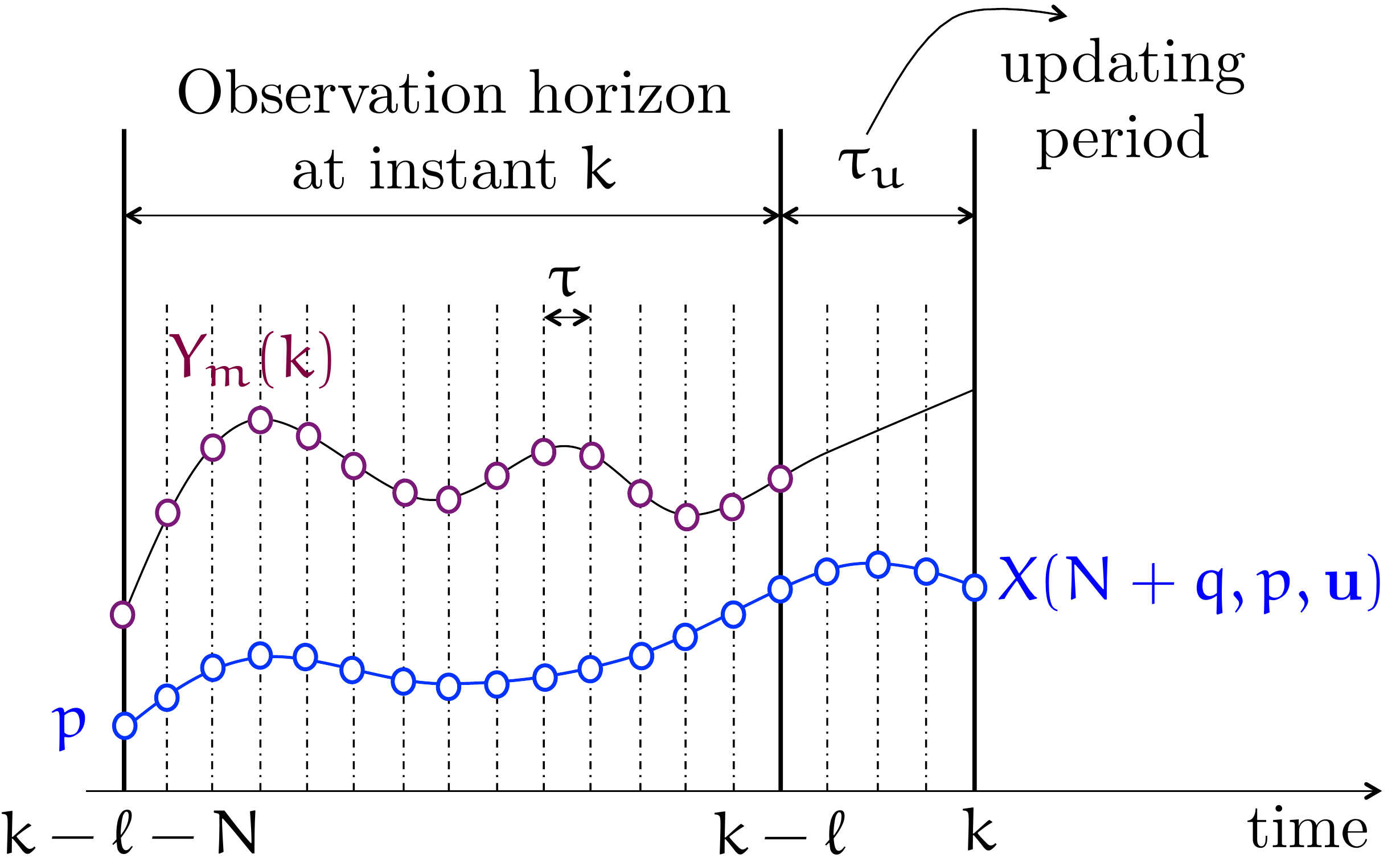}
\caption{Schematic view showing the key notation used in the definition of MHO.} \label{schema_init_q} 
\end{center} 
\end{figure}
Ideal Moving-Horizon Observers ({\bf  MHO})
are algorithms which compute a rational estimation $\hat x(k)$ of the current state $x(k)$ based on the last estimation $p^{opt}(k-\ell)$ of some decision variable $p$ and the sequence of past measured outputs $Y_m(k)$ defined by (see Figure \ref{schema_init_q}):
\begin{eqnarray}
Y_m(k):= \begin{pmatrix}
y_m(k)\cr \vdots\cr y_m(k-N)
\end{pmatrix} \in \mathbb{R}^{(N+1)\cdot n_y} 
\end{eqnarray}
More precisely, the general form of MHO-based state estimation is given by:
\begin{eqnarray}
\hat x(k)&=&F(p^{opt}(k),\bu(k)) \label{xhatgen} \\
p^{opt}(k)&=& \mbox{\rm arg}\min_{p\in \mathbb{P}} J\bigl(p\ \vert\  Y_m(k-\ell),\bu(k-\ell),p^{opt}(k-\ell)\bigr)  \nonumber \\ \label{defdepopt} 
\end{eqnarray} 
where 
\begin{itemize}
\item[$\checkmark$] $p(k-\ell)$ is a vector of decision variables that is generally taken to be the presumed value of the state at the beginning of the observation horizon (at instant $k-\ell-N$) or it can be the whole state trajectory as in the multiple-shooting version of the MHO implementation \citep{Kuhl2011}. \\
\item[$\checkmark$] $\mathbb{P}\subset \mathbb{R}^{n} $ is the set of admissible parameter values.\\
\item[$\checkmark$] $J(\cdot)$ is the cost function that is generally decomposed into an output prediction error-related term and a system's dynamic related term with the standard trade-off weighting coefficients that depend on noise covariance matrices.
\end{itemize} 
\begin{com} \label{pasmonprob} 
\noindent At this stage, it is very important to underline that the definition of the cost function is out of the scope of the present contribution in which the cost function is supposed to be given by the designer. The aim of the forthcoming developments is to suggest a way to distribute the minimization of this cost function over the system lifetime. Whether this enhances a better estimation of the state or not depends on the relevance of the weighting matrices that are used to define this cost function. $\hfill \spadesuit$ 
\end{com}
Figure \ref{schema_init_q} shows a schematic view of the situation when the presumed value of the state at instant $k-\ell-N$ is taken as decision variable, denoted by $p$. Note that in this specific case, the map $F$ involved in (\ref{xhatgen}) is given by:
\begin{eqnarray}
\hat x(k)=X(N+\ell,p^{opt}(k),\bu(k))=:F(p^{opt}(k),\bu(k))
\end{eqnarray} 
In this framework, the time period that lasts between the availability of the measurement data $Y_m(k-\ell)$ (namely instant $k-\ell$) and the delivery of the optimal value $p^{opt}(k)$ (instant $k$) is equal to the so called updating period $\tau_u$ given by:
\begin{eqnarray}
\tau_u=\ell\cdot \tau \label{defdetauu} 
\end{eqnarray} 
During this amount of time, the optimization problem defined by (\ref{defdepopt}) has to be solved. \ \\ \ \\  
For systems that need fast updating periods, the updating time $\tau_u=\ell\tau$ may not be sufficient to reach the optimal solution (regardless of local minima). In such situations, the definition (\ref{defdepopt}) is no more relevant. Indeed, one can only refer to an updating rule that involves a limited number $q$ of iterations of some optimization subroutine $\mathcal S$. This implicitly assumes that $q$ iterations of the subroutine $\mathcal S$ can be performed during the updating period $\ell\tau$. Denoting by $\tau_c$ the time needed to perform a single iteration, it comes that:
\begin{eqnarray}
\ell=\ell(q)=int\Bigl(\dfrac{q\tau_c}{\tau}\Bigr)+1 \label{defdeellq} 
\end{eqnarray} 
where for any positive real $s$, $int(s)$ stands for the integer part of $s$. Based on the above discussion, in the sequel, $\ell$ can be viewed as a function of $q$ for a given pair of measurement acquisition period $\tau$ and an optimization subroutine that defines the time $\tau_c$ on a specific hardware facility.
\ \\
\begin{figure}
\begin{center}
\includegraphics[width=0.5\textwidth]{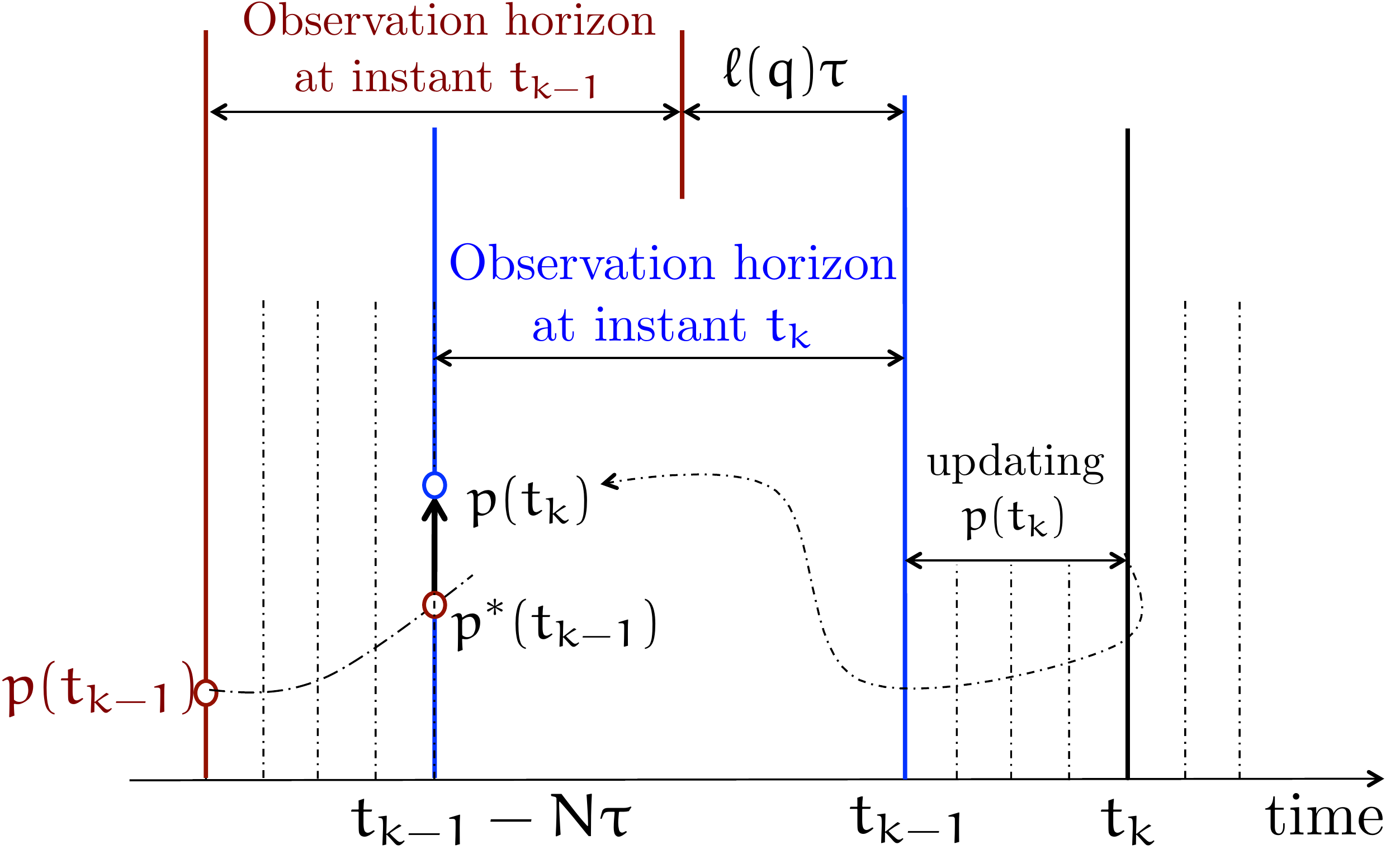}
\caption{The time structure of the updating scheme defined by (\ref{inbvgtrbis1})-(\ref{inbvgtrbis2}).} \label{schema_2_q} 
\end{center} 
\end{figure}
\ \\ 
Consequently, the decision variable $p$ can be updated only at updating instants (see Figure \ref{schema_2_q}):
\begin{eqnarray}
t_k=t_{k-1}+\bigl[\ell(q(t_{k-1}))\bigr]\cdot \tau
\end{eqnarray}  
This leads to the following updating rules: 
\begin{eqnarray}
&&p(t_k)=\mathcal S^{(q)}\bigl(p^*(t_{k})\ \vert Y_m(t_{k-1}),\bu(t_{k-1})\bigr) \label{inbvgtrbis1}\\
&&p^*(t_{k})=X(\ell(t_{k-1}),p(t_{k-1}),\bu(t_{k-1}-N\tau))\label{inbvgtrbis2}
\end{eqnarray} 
Note that the definition of $p^*(t_{k})$ involves an $\ell$-step prediction map. Note that $p^*(t_k)$ is used as an initial guess for the iterations invoked in (\ref{inbvgtrbis1}). \ \\ \ \\ 
Figure \ref{schema_2_q} illustrates the temporal structure of the updating laws given by (\ref{inbvgtrbis1})-(\ref{inbvgtrbis2}). It shows the following features:
\begin{itemize}
\item[$\checkmark$] The shift of the observation horizon corresponds at instant $t_{k}$ to $\ell(q(t_k))$ basic sampling periods leading to a temporal shift of $\ell(q(t_k))\tau$. \\
\item[$\checkmark$] The computation of $p(t_k)$ is done during the time interval $[t_{k-1},t_k]$ based on the initial guess $p^*(t_{k})$.\\ 
\item[$\checkmark$] Note that the updating of the state estimate can still be done at each basic sampling period using the last updated value of $p$ using the following expressions:
\begin{eqnarray}
&& \forall i\in \{1,\dots,\ell\},\label{xhatfinalement}\\
&&\hat x(t_{k-1}+i\tau)=X(\ell+N+i,p(t_{k-1}),\bu(t_{k-1}+i\tau))\nonumber
\end{eqnarray} 
where $\ell=\ell(q(t_{k-1}))$. 
\end{itemize} 
Now given that $p^*(t_{k})$ is defined in terms of $p(t_{k-1})$ and $\bu(t_{k-1}-N\tau)$ [see (\ref{inbvgtrbis2})] and that the latter is contained in $\bu(t_{k-1})$, one can write the  evolution equation for the parameter vector $p$ by combining (\ref{inbvgtrbis1})-(\ref{inbvgtrbis2}) :
\begin{eqnarray}
p(t_k)=G\bigl(p(t_{k-1}),q(t_{k-1}),\bw(t_{k-1})\bigr) \label{tf54de} 
\end{eqnarray} 
for a straightforward definition of $G$ where $\bw(t_{k-1})$ stands for the past measurement data, namely:
\begin{eqnarray}
\bw(t_{k-1}):=\Bigl\{Y_m(t_{k-1}),\bu(t_{k-1})\Bigr\}
\end{eqnarray} 
Note also that using the same notation, the value of the cost function  at instant $t_k$ can be written in the following form [according to (\ref{defdepopt})]:
\begin{eqnarray}
J\bigl(p(t_k)\vert p(t_{k-1}),\bw(t_{k-1})\bigr)
\end{eqnarray} 
and again, since $p(t_k)$ is given by (\ref{tf54de}), the cost function can be written in a more compact form:\begin{eqnarray}
J=h\bigl(p(t_{k-1}),q(t_{k-1}),\bw(t_{k-1})\bigr) \label{defdeh} 
\end{eqnarray} 
\begin{figure}
\begin{center}
\includegraphics[width=0.35\textwidth]{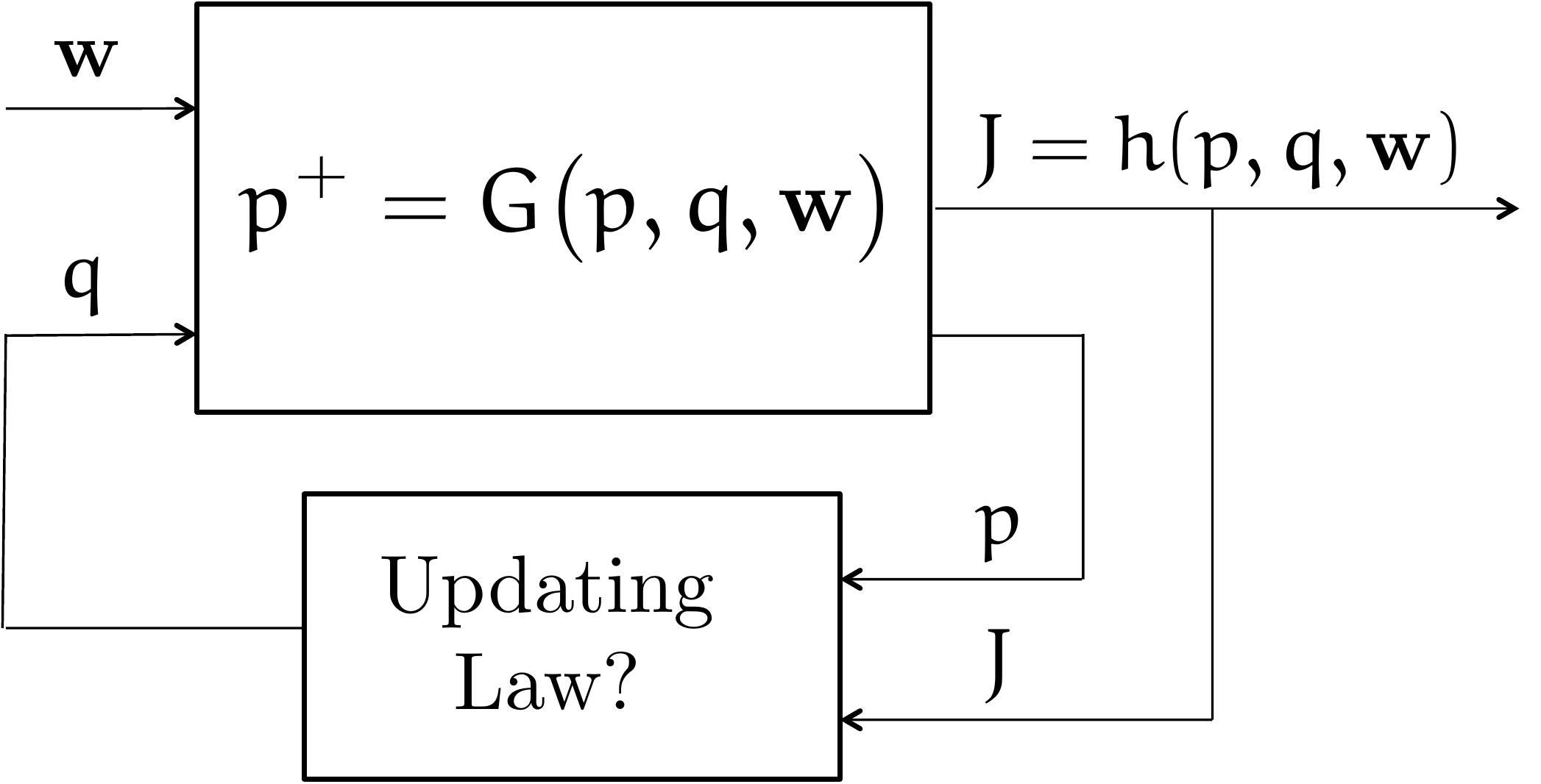}
\caption{The measurement inclusion rate problem viewed as an discrete-time output regulation problem with dynamic state $p$, control input $q$ and a regulated output $J$.} \label{feedback_scheme} 
\end{center} 
\end{figure}
\ \\ The discussion above enables to view the situation as the one in which there is a discrete-time dynamic system defined by (\ref{tf54de}) in which, the state is $p$, the control is given by $q$ and the past measurement (including the control and the measured output) are exogenous non modeled signals while the control objective is to steer the output $J$ given by (\ref{defdeh}) to its minimum value. \ \\ \ \\ 
This is obviously a control problem (see Figure \ref{feedback_scheme}) in which the control variable is defined by $q$ which is the number of measurements that enter (and leave) the buffer of the MHO at the next shift of the observation interval. \ \\ \ \\ 
Solving this control problem leads to an adaptive behavior of $q$ that takes into account both the measurement buffer and the measured behavior of the cost function during the system lifetime. The decision variable $q$ is called hereafter the {\em Measurement Inclusion Rate} as it defines the updating time $\ell(q)\tau$ during which no new measurements are accounted for and iterations are applied to the same cost function. 
\begin{com} \label{comentairetemps} 
Note that the computations involved in the updating rule for $q$ (the feedback law) must correspond to a negligible burden since the whole framework is supposed to compensate for the lack of computation time (see section \ref{seccomplexity}). 
\end{com}
\section{Updating Law For The Measurement Inclusion Rate}\label{secupdatinglaw} 
\noindent In order to simplify the expressions, the following short notation is used:
\begin{itemize}
\item[$\checkmark$] $J(t_k)$ denotes the best obtained value of the cost function defined on the observation horizon $[t_{k-1}-N\tau,t_{k-1}]$ after $q$ iterations. \\
\item[$\checkmark$] $J^*(t_k)$ denotes the value of the same cost function at the initial guess $p^*(t_{k})$ that is computed according to (\ref{inbvgtrbis2}). 
\end{itemize} 
Obviously, one key step toward understanding the convergence issue is to examine the ratio between the best obtained values of the cost function at two successive updating instants $t_{k-1}$ and $t_k$, namely (see Figure \ref{schema_2_q}):
\begin{eqnarray}
\dfrac{J(t_k)}{J(t_{k-1})}&=&\underbrace{\dfrac{J(t_k)}{J^*(t_k)}}_{E(q,t_k)}\times \underbrace{\dfrac{J^*(t_k)}{J(t_{k-1})}}_{D(q,t_k)} \label{defdeDE}\\
&=& E(q,t_k)\times D(q,t_k)=:K(q,t_k) \label{defdeKKKK} 
\end{eqnarray} 
Note however that in order for the ratios involved in (\ref{defdeDE}) to be well defined, the following easy-to-meet assumption is needed: 
\begin{ass}
There is a positive $c>0$ such that for all $t$ and all $p$, one has: $J(p,t)\ge c$. $\hfill \spadesuit$
\end{ass}
Note that this can be fulfilled by adding $c$ to any original nonnegative cost function's definition. \ \\ \ \\ 
The terms $E(q,t_k)$ and $D(q,t_k)$ are similar to the terms invoked in \citep{Pavia2008,AlamirECC2013} where the dual MPC problem is studied. More precisely:
\begin{itemize}
\item[$\checkmark$] $E(q,t_k)$ is linked to the {\em local} efficiency of the optimizer's iterations as it represents the contraction of the cost function due to the execution of $q$ successive iterations. This ratio is obviously lower or equal to $1$. \\
\item[$\checkmark$] $D(q,t_k)$ is the ratio between the value of the cost function after horizon shift and using the model-based predicted value $p^*(t_{k})$ that is compatible with the last achieved value $p(t_{k-1})$ (hot start). In the case where $p(t_{k-1})$ matches the true value of the state $x(t_{k-1}-N\tau)$ then in the ideal case (no model discrepancies and no measurement noise), the ratio $D(q,t_k)$ is equal to $1$.  Consequently, when this is not the case, $D(q,t_k)$ can be viewed as a {\em disturbance indicator} which gathers all the unavoidable above mentioned imperfections.  Note also that since the length of the updating period $\tau_u=\ell\tau=int(q\tau_c/\tau)+1$ depends on $q$, the disturbances induced ratio $D(q,t_k)$ depends also on $q$. 
\end{itemize} 
Based on the discussion above, the following model of the disturbance ratio is used in the sequel:
\begin{eqnarray}
D(q,t_k)=1+\alpha(t_k)\cdot q \label{structuredeD} 
\end{eqnarray} 
where $\alpha(t_k)$ is a parameter to be identified on-line as shown later. \ \\ \ \\ 
Now, given the evolution equation:
\begin{eqnarray}
J(t_k)=K(q,t_k)\cdot J(t_{k-1}) \label{defdeKKK} 
\end{eqnarray} 
it becomes obvious that one rationale that can be used in the derivation of the updating law for $q$ is to force the multiplicative gain $K$ to be lower than one as this enhances the convergence of the cost function $J$ and when this goal is achievable, $q$ must be monitored so that the response time of the {\em closed-loop} continuous-time system (in the sense of Figure \ref{feedback_scheme}) is minimized.   \ \\ \ \\ 
These considerations lead to the following ideal updating law for $q$:
\begin{eqnarray}
&&q(t_{k+1})=\nonumber \\
&&arg\min_{q\in \{q_{min},\dots,q_{max}\}} \left\{ 
\begin{array}{ll}
 \dfrac{q}{\vert \log(K(q,t_k))\vert}& \mbox{\rm if $K(q(t_k),t_k)<1$}\\ &\ \\
K(q,t_k) & \mbox{otherwise}
\end{array}
\right. \label{idealoptim} 
\end{eqnarray} 
This is because the ratio:
$$\dfrac{q}{\vert \log(K(q,t_k))\vert}$$
\ \\
is almost proportional (up to the {\em int} function discontinuity) to the response time of a discrete-time dynamics (\ref{defdeKKK}) that is  characterized by the discrete pole $K$ and the sampling time $\ell(q)\tau$ where $\ell(q)$ is given by (\ref{defdeellq}). Note that $q_{min}\ge 2$ is systematically considered in order to be always capable of estimating the gradient of the cost function involved in (\ref{idealoptim}) w.r.t $q$ as it is explained in the sequel. \ \\ \ \\ 
The exact solution of the optimization problem (\ref{idealoptim}) would needs too many computations that would make the updating rule inappropriate (see Remark \ref{comentairetemps}). Instead, an approximated gradient approach is used following the ideas proposed in \citep{AlamirECC2013} in the case of MPC framework. \ \\ \ \\ 
To do this, the  sensitivity of $K(\cdot,t_k)$ w.r.t $q$ is computed by computing those of $E(\cdot,t_k)$ and $D(\cdot,t_k)$ using the available algorithm data at the past updating instant. Then a quantized gradient step is applied to update the value of $q(t_{k+1})$. This is detailed in the following section. 
\subsection{Updating Algorithm}
\noindent Note first of all that since the past value $q(t_{k})\ge 2$, it is possible to compute the following approximation of the gradient of the efficiency map:
\begin{eqnarray}
\dfrac{\Delta E}{\Delta q}(t_k)\approx \dfrac{J(p^{(q(t_k))},t_{k-1})-J(p^{(q(t_k)-1)},t_{k-1})}{J(p^*(t_{k}),t_{k-1})} \label{dEdq} 
\end{eqnarray} 
On the other hand, using the presumed structure (\ref{structuredeD}) of $D$, one can use the available algorithm's data to compute an estimation of $\alpha(t_k)$ which is nothing but the gradient of $D$ w.r.t $q$, that is:
\begin{eqnarray}
\dfrac{\Delta D}{\Delta q}(t_k)\approx \dfrac{1}{q(t_k)}\Bigl[\dfrac{J^*(t_k)}{J(t_{k-1})}-1\Bigr]\approx \alpha(t_k) \label{dDdq} 
\end{eqnarray} 
Now, using equations (\ref{dEdq})-(\ref{dDdq}), the gradient of the multiplicative gain $K$ w.r.t $q$ can be computed according to:
\begin{eqnarray}
\dfrac{\Delta K}{\Delta q}(t_k)\approx E(t_k)\dfrac{\Delta D}{\Delta q}(t_k)+D(t_k)\dfrac{\Delta E}{\Delta q}(t_k) \label{gradK} 
\end{eqnarray} 
and having this estimation of the gradient, the gradient of the cost function involved in the ideal updating rule (\ref{idealoptim}) can be  computed:
\begin{eqnarray}
&&\dfrac{\Delta\bigl(q/\vert \log(K(t_k))\vert\bigr)}{\Delta q}\approx \dfrac{-\log(K(t_k))+\dfrac{q}{K(t_k)}\times \dfrac{\Delta K(t_k)}{\Delta q}}{\bigl[\log(K(t_k))\bigr]^2}\nonumber \\
&& \label{defdestgfrde} 
\end{eqnarray} 
Using the above computed quantities, the following algorithm can be used to compute the updated value $q(t_{k+1})$:
 \begin{algorithm}[H]  
\caption{Updating rule $q(t_{k+1})=U(q(t_k),t_{k})$}
\label{algupdateq}     
\begin{algorithmic}[1] 
\State{{\bf If} ($K(t_k)\ge 1$) {\bf then}}
\State{$\quad$ $\Gamma\leftarrow \dfrac{\Delta K}{\Delta q}(q(t_k))$ $\quad$ [see (\ref{dEdq})-(\ref{dDdq}) and (\ref{gradK})]}
\State{{\bf Else} }
\State{$\quad$ $\Gamma\leftarrow \dfrac{\Delta (q/\vert \log(K(t_k))\vert)}{\Delta q}$} $\quad$ [see (\ref{defdestgfrde})]
\State{{\bf End If}}
\State{$q(t_{k+1})\leftarrow \max\Bigl\{q_{min},\min\bigl\{q_{max},q(t_k)-\delta\cdot sign(\Gamma)\bigr\}\Bigr\}$}
\end{algorithmic}
\end{algorithm}
In this algorithm, $\Gamma$ represents the gradient of the quantity to be minimized. A quantized step ($\delta \in \mathbb{N}$) in the opposite direction is implemented in Step 6 to update the value of $q$. 
\subsection{Complexity Analysis} \label{seccomplexity} 
Based on the expressions (\ref{dEdq})-(\ref{defdestgfrde}), one can construct Table \ref{tablecalculs} that shows the additional number of elementary operations that are needed to compute the updating law involved in Algorithm \ref{algupdateq}. By {\em additional}, it is meant that the computation of the cost function $J(t_k)$, $J^*(t_{k-1})$ and $J(t_{k-1})$) are excluded because they are by-products of the optimization process and are therefore computed regardless of whether an updating process is used or not.  \\
\begin{table}[H]
\begin{center}
\begin{tabular}{|l|l|l|l|l|l|}
 \hline
Expression& Equation & ($\pm$) &$(\times)$ & $(\div)$ & $\log$\\ \hline
$K$ & (\ref{defdeKKKK})& \  & $2$ & $1$ \    & \  
\\ 
$(\Delta E)/(\Delta q)$ & (\ref{dEdq})& $1$ &\ & $1$ \    & \  \
\\ 
$(\Delta D)/(\Delta q)$ & (\ref{dDdq})& $1$ &\ & $2$ \    & \  \ \\
$(\Delta K)/(\Delta q)$ & (\ref{gradK})& $3$ & $2$ & $3$ \    & \  \\ 
$\Gamma$ & (\ref{defdestgfrde})& $4$ & $5$ & $6$ \    & $1$  \\ \hline 
{\bf Algorithm} \ref{algupdateq}  &  \ & $\mathbf{5}$ & $\mathbf{5}$ & $\mathbf{6}$ \    & $\mathbf{1}$  \\
\hline
\end{tabular}
\end{center}
\vskip 0.2cm       
\caption{Elementary computations involved in the updating law $q(t_{k+1})=U(q(t_k),t_{k})$.}\label{tablecalculs}
\end{table}
\noindent Table \ref{tablecalculs} clearly shows that the computations involved in the updating rule involves only few arithmetic operations and a single logarithm computation). 
\section{Illustrative example} \label{secillust} 
\noindent Let us consider the following modified van-der-pol system:
\begin{eqnarray*}
\dot{x}_1&=&x_2 \\
\dot x_2&=&-ax_1+(1-ux_3x_1^2)x_2\\
\dot x_3&=&0
\end{eqnarray*} 
The measured output is defined by:
\begin{eqnarray*}
y=x_1+\nu
\end{eqnarray*}
where $\nu$ is a white noise.  The parameter $a$ is a parameter that can be badly known in order to enhance the uncertainty feature of the model. The basic sampling (measurement acquisition) period is taken equal to $\tau=2\ ms$. \ \\ \ \\ 
The optimization subroutine involved in (\ref{inbvgtrbis1}) is based on a fast gradient approach \citep{Nesterov1983,Nesterov2004} with restart mechanism \citep{Odonoghue2012}. The details of these algorithm can be found also in \citep{AlamirECC2013}. The explicit definition of these algorithm is not mandatory here since the proposed approach is generic and the use of fast gradient approach as an instantiation of $\mathcal{S}$ is only a matter of choice that enables the main idea to be illustrated. Note that the optimization is performed assuming the following box constraints that are to be interpreted component-wise:
\begin{eqnarray}
\begin{pmatrix}
-10\cr -10\cr 0.1
\end{pmatrix} \le x\le \begin{pmatrix}
10\cr 10\cr 40
\end{pmatrix} 
\end{eqnarray}  \ \\ \ \\ 
The computation time $\tau_c$ needed to perform a single iteration of the subroutine $\mathcal S$ [see (\ref{defdeellq})] is given by $\tau_c=500\ \mu sec$  
The minimum and the maximum number of iterations involved in the updating rule appearing in the step 6. of  Algorithm \ref{algupdateq} are $q_{min}=20$ and $q_{max}=1000$. \ \\ \ \\ 
The observation horizon is taken equal to $N=200$ basic sampling period leading to a time window of $T=0.4\ sec$. 
The variance of the measurement noise is taken equal to $\nu=0.03$. The cost function $J(p)$ that has been used in all the simulations takes the following form at instant $k$:
\begin{eqnarray*}
J:=\sum_{i=0}^{N} \|\hat{y}(k+i\vert p)-y(k+i)\|^2+\rho \|p-\hat{p})\|^2
\end{eqnarray*} 
where $\hat{p}$ is the estimated value based on the past estimation of $p$. The coefficient $\rho=0.01$ is used in the sequel. \ \\ \ \\
In the following sections, the validation scenario is clearly stated and the comparison indicators are defined. 
\subsection{Validation Scenarios}
\noindent For each comparison, $N_s:=50$ scenarios are executed using different values of the initial estimated state $\hat x_0$. These values are randomly chosen according to:
\begin{eqnarray}
\hat x^0:=(0.2\times \mathbb{I}+1.8 \begin{pmatrix}
r_1&0&0\cr 0&r_2&0\cr 0&0&r_3
\end{pmatrix})x^0 \label{tgf65} 
\end{eqnarray} 
where $x^0=(3,1,1)^T$ is the true initial state. $r_i$ are uniformly distributed random variable belonging to the interval $[0,1]$. Note that (\ref{tgf65}) simply means that each component $\hat x_i^0$ of the initial state of the observer is randomly chosen in the interval $[0.2x_i^0,2x_i^0]$. The resulting set of observer initial states is denoted hereafter by $\hat{\mathbb{X}}^0$.\ \\ \ \\ 
Simulations are performed during $N_{sim}=2000$ sampling periods ($4\ sec$) using the following input profile: 
$$u(t):=1-\dfrac{1}{2}cos(2t)$$
Five observer settings are compared which are:
\begin{enumerate}
\item {\bf Setting 1}.  $q=q_{min}=20$. No updating is used. 
\item {\bf Setting 2}.  $q=q_{min}=50$. No updating is used. 
\item {\bf Setting 3}.  $q=q_{min}=100$. No updating is used. 
\item {\bf Setting 4}.  $q=q_{min}=300$. No updating is used. 
\item {\bf Setting 4}.  $q(0)=q_{min}=20$. Updating is used for $q$ using the rule expressed in Algorithm \ref{algupdateq} starting from the initial value $q(0)=q_{min}=20$ and the increment size $\delta=10$. Note that the first four setting can be obtained using $\delta=0$ while initializing $q$ to $20$, $50$, $100$ and $300$ respectively. 
\end{enumerate} 
Note that for each of the above mentioned five settings, the same $50$ trials leading to the set $\hat{\mathbb{X}}
^0$ are used. That is, the trials are first done and then the $50$ simulated scenarios are executed for each of the five settings leading to $250$ simulations. This avoid biased comparison that may be due to different trials for each setting. 
\subsection{Performance indicator}
\begin{figure}
\begin{center}
\includegraphics[width=0.42\textwidth]{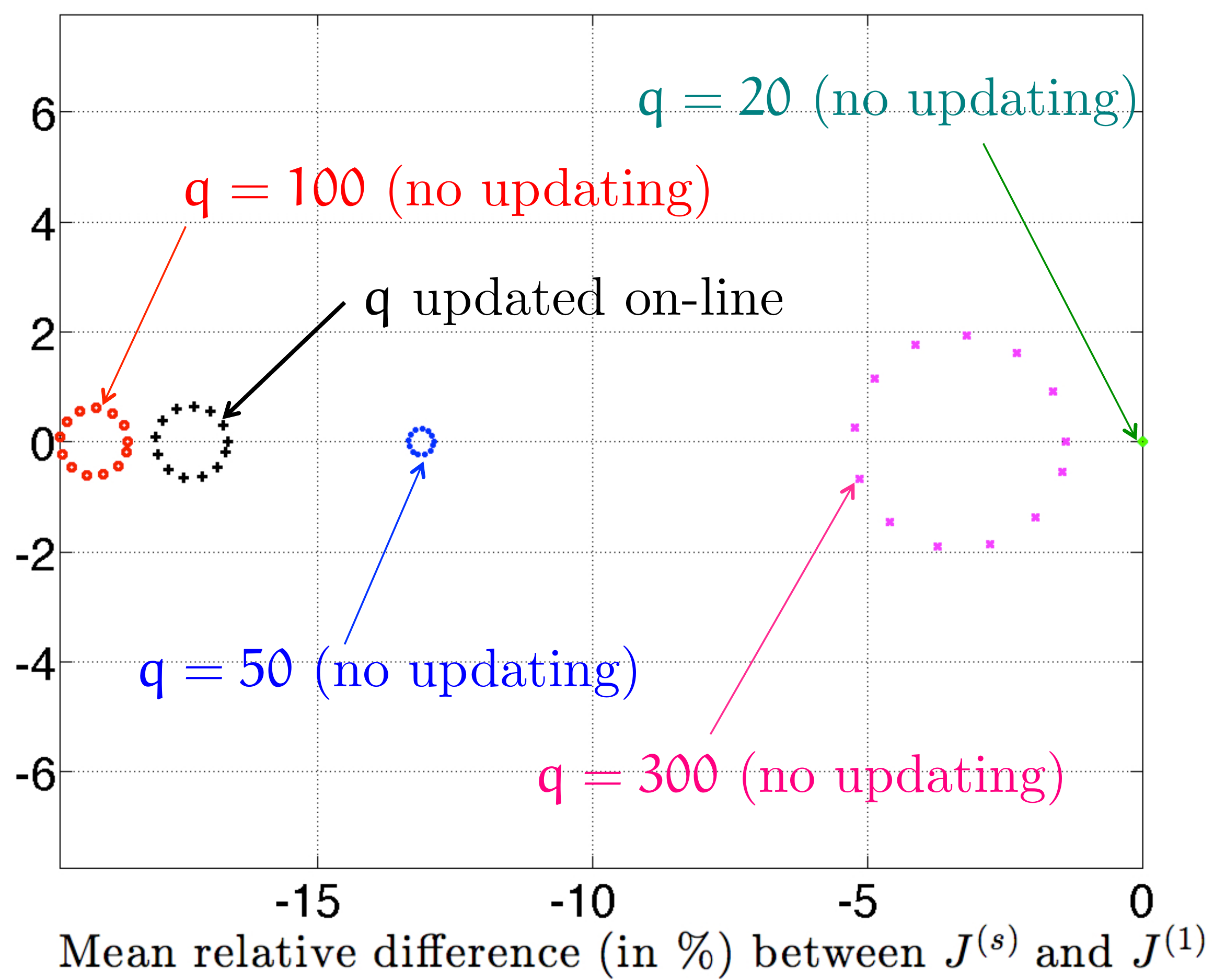}\ \\
(a)\ \\
\includegraphics[width=0.42\textwidth]{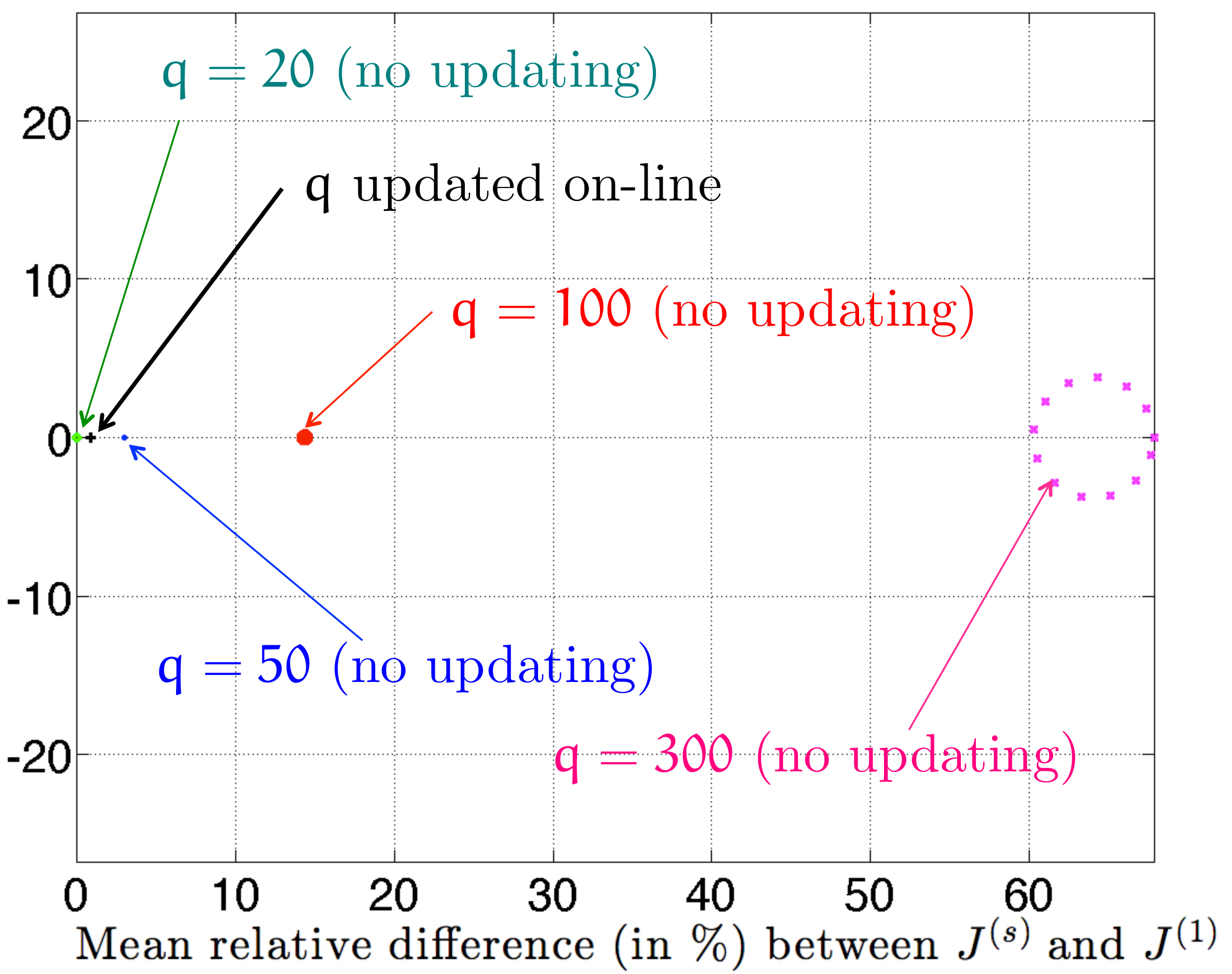}\ \\
(b)
\end{center} 
\caption{\color{Blue} Comparison of the performance of the observer under different constant number of iterations $q\in \{20,50,100,300\}$ on one side and under the proposed updating scheme that start at $q(0)=20$. {\bf (a)} Case where the observer knows the exact value of the parameter $a=10$. {\bf (b)} Case where the observer uses an erroneous value $\hat a=7$ instead of $a=10$. The radius of each circle is equal to twice the corresponding variance.} \label{sim1-2} 
\end{figure}

\noindent As mentioned in Remark \ref{pasmonprob}, we are interested in the behavior of the cost function. That is the reason why the characteristics (mean and variance) of the following quantities are monitored:
$$\hat{J}^{(s)}(k\vert \hat x^0)\ \mbox{\rm for}\  k=1,\dots,N_{sim}-N\ \mbox{\rm and}\  \hat x^0\in \hat{\mathbb{X}}^0
$$ 
where $\hat J^{(s)}(k\vert \hat x^0)$ is the cost function at instant $k$ for the scenario starting at $\hat x^0\in \hat{\mathbb{X}}^0$ and using the setting number $s$.\ \\ \ \\ 
More precisely, for easiness of comparison, the following quantities are considered for all setting index $s$:
\begin{eqnarray}
m^{(s)}&:=& \displaystyle{{\rm Mean}_{k,\hat x^0}} \Bigl[\dfrac{J^{(s)}(k\vert \hat x^0)-J^{(1)}(k\vert \hat x^0)}{J^{(1)}(k\vert \hat x^0)}\Bigr] \label{defdems}\\
 \sigma^{(s)}&:=& \displaystyle{{\rm Var}_{k,\hat x^0}} \Bigl[\dfrac{J^{(s)}(k\vert \hat x^0)-J^{(1)}(k\vert \hat x^0)}{J^{(1)}(k\vert \hat x^0)}\Bigr] \label{defdess}
\end{eqnarray} 
where the mean and the variance invoked in (\ref{defdems}) and (\ref{defdess}) are taken on the set of values of $(k,\hat x^0)$ given by: 
$$\Bigl\{1,\dots,N_{sim}-N\Bigr\}\times \hat{\mathbb{X}}^0$$
Two scenarios are used to assess the relevance of the updating rule. In the first the parameter $a=10$ is perfectly known by the observer while in the second scenario, the observer uses an erroneous value $\hat a=7$. Figure \ref{sim1-2} shows the corresponding performance of the different observers using circles that are centered at $m^{(s)}$, $s=1,\dots,5$ and with radius that are equal to twice the corresponding variance $\sigma^{(s)}$.  In the first case, an optimal updating period seems to be close to $q=100$ (see Figure \ref{sim1-2}.a). However, when model uncertainties increase (as in the second scenario), the optimal updating period becomes the minimum value $q=q_{min}$. Note that in both cases, the performance of the updated version systematically spots close to the optimal performance. Note also that if the parameter $a$ changes during the scenario, then obviously the observer with the proposed updating scheme will be better that any observer with constant updating period.  
\section{Conclusion and Future Work} \label{secconc} 
\noindent In this paper, a novel updating rule for the measurement inclusion rate in MHO is proposed and validated through a simple example. The proposed rule enables near optimal performance to be achieved in presence of unavoidable, unpredictable model discrepancies and without off-line extended tuning. Current investigation focuses on the validation of the proposed methodology on real-world estimation problems. 
\bibliographystyle{plain}
\bibliography{bibobs}
\end{document}